\documentclass{elsart}
\usepackage{amsfonts}
\usepackage{amsmath}
\usepackage{amssymb}
\usepackage{graphicx}
\usepackage{float}
\usepackage{lmodern}
\usepackage{amssymb}
\newtheorem{proposition}{Proposition}
\newtheorem{remark}{Remark}

\newtheorem{theorem}{Theorem}

\newtheorem{corollary}{Corollary}
\numberwithin{equation}{section}
\theoremstyle{plain}
\begin{document}

%TCIMACRO{\TeXButton{Begin frontmatter}{\begin{frontmatter}}}%
%BeginExpansion
\begin{frontmatter}%
%EndExpansion

%TCIMACRO{\TeXButton{Title}{\title{Sample Elsevier Document}}}%
%BeginExpansion
\title{Discrete coherent states for higher Landau levels}%
%EndExpansion

%TCIMACRO{\TeXButton{Author}{\author{A. U. Thor}}}%
%BeginExpansion
\author{L. D. Abreu}%
%EndExpansion

%TCIMACRO{\TeXButton{Address}{\address{Institute of Far Away Places}}}%
%BeginExpansion
\address{ARI, Austrian Academy of Sciences, Wohllebengasse 12-14, A-1040, Vienna, Austria; e-mail: labreu@kfs.oeaw.ac.at}%
%EndExpansion

%TCIMACRO{\TeXButton{Address}{\address{Institute of No Far Away Places}}}%
%BeginExpansion

%EndExpansion

%TCIMACRO{\TeXButton{Collaborator}{\collab{Me Too}}}%
%BeginExpansion
\collab{P. Balazs}%
%EndExpansion

%TCIMACRO{\TeXButton{Address}{\address{Institute of Far Away Places}}}%
%BeginExpansion
\address{ARI, Austrian Academy of Sciences, Wohllebengasse 12-14, A-1040, Vienna, Austria; e-mail: peter.balazs@oeaw.ac.at}%
%EndExpansion

%TCIMACRO{\TeXButton{Collaborator}{\collab{Me Too}}}%
%BeginExpansion
\collab{M. de Gosson}%
%EndExpansion

%TCIMACRO{\TeXButton{Address}{\address{Institute of Far Away Places}}}%
%BeginExpansion
\address{ NuHAG, University of Vienna, Oskar-Morgenstern-Platz 1 A-1090 Vienna, Austria; e-mail: maurice.degosson@gmail.com }%
%EndExpansion

%TCIMACRO{\TeXButton{Collaborator}{\collab{Me Too}}}%
%BeginExpansion
\collab{Z. Mouayn}%
%EndExpansion

%TCIMACRO{\TeXButton{Address}{\address{Institute of Far Away Places}}}%
%BeginExpansion
\address{Department of Mathematics, Faculty of Sciences and Technics
(M'Ghila) \\ P.O. Box 523, Beni Mellal, Morocco; e-mail: mouayn@gmail.com}%
%EndExpansion

%TCIMACRO{\TeXButton{Begin abstract}{\begin{abstract}} }%
%BeginExpansion
\begin{abstract}
%EndExpansion
We consider the quantum dynamics of a charged particle evolving under the
action of a constant homogeneous magnetic field, with emphasis on the
discrete subgroups of the Heisenberg group (in the Euclidean case) and of
the $SL(2,\mathbb{R})$ group (in the Hyperbolic case).\ We investigate
completeness properties of discrete coherent states associated with higher
order Euclidean and hyperbolic Landau levels, partially extending classic results of
Perelomov and of Bargmann, Butera, Girardello and Klauder. In the Euclidean
case, our results follow from identifying the completeness problem with
known results from the theory of Gabor frames. The results for the hyperbolic
setting follow by using a combination of methods from coherent states,
time-scale analysis and the theory of Fuchsian groups and their
associated automorphic forms.
%TCIMACRO{\TeXButton{End abstract}{\end{abstract}}}%
%BeginExpansion
\end{abstract}%
%EndExpansion

%TCIMACRO{\TeXButton{Begin keywords}{\begin{keyword}} }%
%BeginExpansion
%EndExpansion

%TCIMACRO{\TeXButton{End frontmatter}{\end{frontmatter}}}%
%BeginExpansion
\end{frontmatter}%
%EndExpansion

\section{Introduction}

In this paper we consider the quantum dynamics of a charged particle
evolving under the action of a constant homogeneous magnetic field, first in
the Euclidean and then in the hyperbolic setting. The goal is to construct
discrete coherent states associated with the evolution of the particle when
higher Landau levels are formed and to obtain conditions on the completeness
of such coherent states. This extends well known results of Perelomov \cite%
{Perelomov} and of Bargmann, Butera, Girardello and Klauder \cite{BargmannEt}%
. In the first part of the paper, we consider a constant magnetic field
acting on the Euclidean space realized as the complex plane $\mathbb{C}$,
leading to the formation of a discrete spectrum known as the Euclidean
Landau Levels. In the second part of the paper, we let a constant magnetic
field act on the\ open hyperbolic plane realized as the Poincar\'{e} upper
half-plane $\mathbb{C}^{+}=\{z\in \mathbb{C},\Im z>0\}$, leading to the
formation of a mixed spectrum, with a \emph{discrete part} corresponding to
\textit{bound states }(\emph{hyperbolic Landau levels}) and a continuous
part corresponding to \textit{scattering states}.

The concept of a set of states on a lattice in phase space was first
considered by J. von Neumann in the Euclidean case \cite{vonNeumann}. It
became physically very attractive because it contains the fundamental
commutation relations of quantum mechanics. Indeed, lattices have an
underlying unit cell (fundamental domain) related to the size of the Plank
constant.

\begin{center}
\begin{figure}[H]
    \centering
        \includegraphics[width=0.60\textwidth]{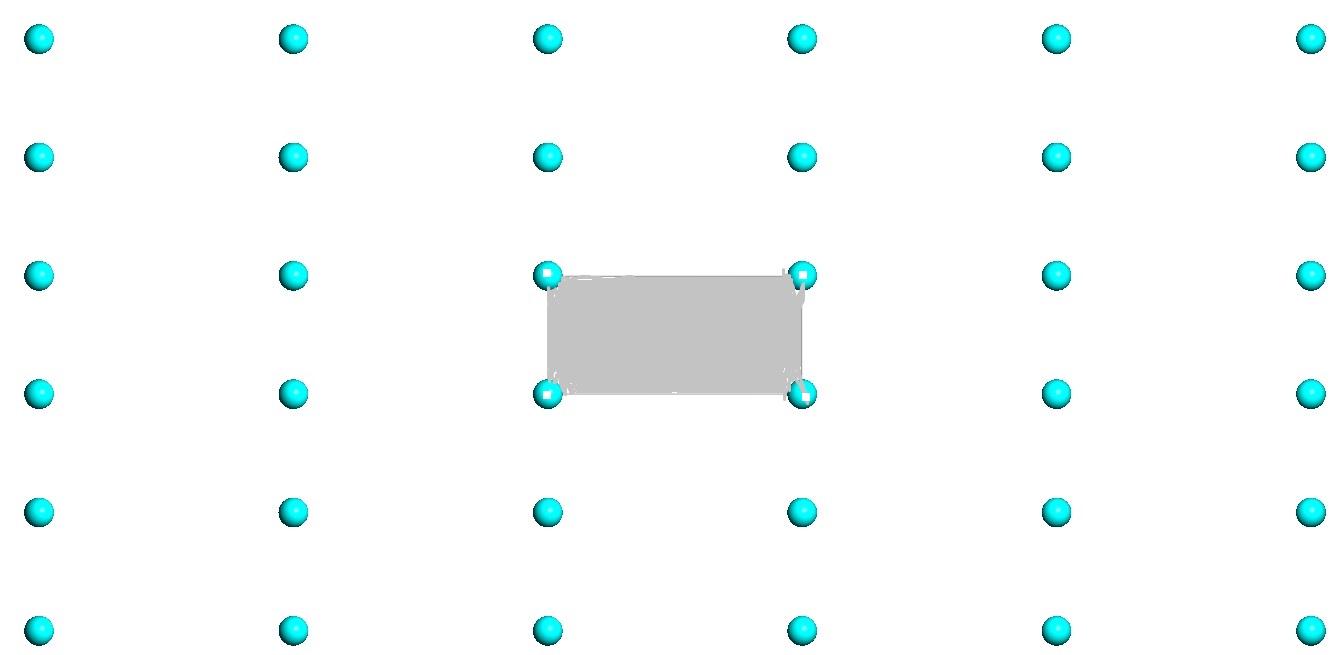}		
		\
		\caption{\small{Picture 1. An Euclidean lattice and a fundamental domain. See section 2.3
		 }}
\end{figure}
\end{center}

In his treatment of quantum mechanics \cite{vonNeumann}, J. von Neumann
raised the question of completeness of coherent states indexed by a lattice.
The question turned out to be nontrivial from a mathematical point of view
and, so far, it has only been fully understood for some special coherent
states. This is the case of the coherent states associated with the first
Landau Level. The situation has been clarified in \cite{BargmannEt} and \cite%
{Perelomov}, because it can be related to the structure of zeros of analytic
functions, where classical methods from complex analysis can be used.
However, in higher Landau Levels, even the case of the Euclidean Landau
levels is not yet fully understood. In both the Euclidean and Hyperbolic
setting, one has to deal with spaces of polyanalytic functions \cite%
{Abreusampling}, \cite{HendHaimi}, \cite{SuperWave}. Since polyanalytic
functions have a much more complicated structure of zeros \cite{Balk},
several essential tools from complex analysis cannot be applied. However, in
recent years, important progress has been made by combining analytic
function theory with methods from time-frequency analysis \cite{CharlyYura},
\cite{Abreusampling}, \cite{Bracken}. The purpose of the first part of this
paper is to translate these results from time-frequency analysis to the
setting of coherent states attached to higher Landau Levels. This has a
twofold purpose: to bring the results to the attention of the physics
community and to motivate the results on the hyperbolic setting of the
subsequent section, where time-scale (wavelet) theory replaces
time-frequency (Gabor) analysis. Indeed, our main object of study in the
paper is the quantum dynamics of a charged particle evolving on the open
hyperbolic plane under the action of a constant magnetic field. While
previous work on this problem has been concerned with the spectral
properties of the corresponding Landau Hamiltonian \cite{Comtet}, \cite%
{Groshe} and their associated continuous coherent states \cite{Mouayn}, the
investigation of the associated discrete coherent states labeled by discrete
subgroups of $PSL(2,\mathbb{R})=SL(2,\mathbb{R})/\{\pm I\}$ seems to have
been overlooked. The discrete coherent states are relevant for the
understanding the hyperbolic setting because the nontrivial dynamics is
induced by the tesselation of the Poincar\'{e} plane by discrete subgroups
of $PSL(2,\mathbb{R})$, which are called \emph{Fuchsian groups}. Important
examples of Fuchsian groups are provided by the modular group $PSL(2,\mathbb{%
Z})$ and by the congruence groups of order $n$. Some background and examples
of Fuchsian groups are given in the last section. This is a remarkable
instance of the usefulness of analytic number theory in a physical problem.
The idea of using Fuchsian groups as a replacement for the Euclidean
lattices seems to have first been used by Perelomov, who provides a full
analysis of the first hyperbolic Landau level in \cite[Chapter 14]%
{Perelomovbook}, where the analysis is done in the disc. In the present
paper we make the corresponding analysis for the higher hyperbolic Landau
levels. As the unit cell of the model one considers a fundamental domain for
the group. For instance, the set%
\begin{equation*}
D=\left\{ z\in \mathbb{C}^{+}:\left\vert z\right\vert \geq 1\text{ and }%
\left\vert \Re z\right\vert \leq \frac{1}{2}\right\}
\end{equation*}%
is a fundamental domain for the modular group $PSL(2,\mathbb{R})$.

The shadow area in the next image represents the fundamental domain $D$.

\begin{center}
\begin{figure}[H]
    \centering
        \includegraphics[width=0.60\textwidth]{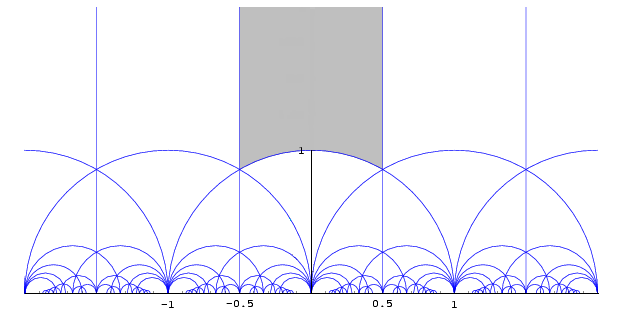}
				\
		\caption{\small{Picture 2. The modular group $PSL(2,\mathbb{Z})$. See section 3.5
		 }}
\end{figure}
\end{center}
The following terminology will be used. A functional Hilbert space $\mathcal{%
H}$ has a system $\{f_{g}\}$ of coherent states, labelled by elements $g$ of
a locally compact group $G$ if:

$\left( i\right) $ There is a representation $T:g\rightarrow T_{g}$ of $G$
labelled by unitary operators $T_{g}$ on $\mathcal{H}$

$\left( ii\right) $ There is a vector $f_{0}\in \mathcal{H}$ such that for $%
f_{g}=T_{g}\left[ f_{0}\right] $ and for arbitrary $f\in \mathcal{H}$ we
have:
\begin{equation*}
\left\langle f,f\right\rangle _{\mathcal{H}}=\int_{G}\left\vert \left\langle
f,f_{g}\right\rangle \right\vert ^{2}d\nu (g)\text{,}
\end{equation*}%
where $d\nu $ stands for the left Haar measure of $G$.

The core of the paper is organized in two sections and an appendix with the more
technical proofs. Section 2 deals with Euclidean Landau levels and section 3
with their hyperbolic analogues. In each of the sections, after providing
some background on the mathematical and physical model, we first construct
the coherent states associated with the higher levels and then investigate
their discrete counterparts. We finish with a short conclusion including some remarks about the theoretical methodology, highlighting the interaction between physical and signal analysis which has made possible the investigations carried out in this paper.

\section{ Euclidean Landau levels}

\subsection{Definitions}

The Hamiltonian operator describing the dynamics of a particle of charge $e$
and mass $m_{\ast }$ on the Euclidean $xy$-plane, while interacting with a
perpendicular constant homogeneous magnetic field, is given by the operator
\begin{equation}
H:=\frac{1}{2m_{\ast }}\left( i\hbar \nabla -\frac{e}{c}\mathbf{A}\right)
^{2}\text{,}  \label{2.1.1}
\end{equation}%
where $\hbar $ denotes Plank's constant, $c$ is the light speed and $i$
the imaginary unit. Denote by $B>0$ the strength of the magnetic field and
select the symmetric gauge%
\begin{equation*}
\mathbf{A=-}\frac{\mathbf{r}}{2}\times \mathbf{B=}\left( -\frac{B}{2}y,\frac{%
B}{2}x\right) \text{,}
\end{equation*}%
where $\mathbf{r}=\left( x,y\right) \in \mathbb{R}^{2}$. For simplicity, we
set $m_{\ast }=e=c=\hbar =1$ in (\ref{2.1.1}), leading to the Landau
Hamiltonian
\begin{equation}
H_{B}^{L}:=\frac{1}{2}\left( \left( i\frac{\partial }{\partial x}-\frac{B}{2}%
y\right) ^{2}+\left( i\frac{\partial }{\partial y}+\frac{B}{2}x\right)
^{2}\right)  \label{2.1.3}
\end{equation}%
acting on the Hilbert space $L^{2}\left( \mathbb{R}^{2},dxdy\right) $. The
spectrum of the Hamiltonian $H_{B}^{L}$ consists of infinite number of
eigenvalues with infinite multiplicity of the form
\begin{equation}
\epsilon _{n}^{B}=\left( n+\frac{1}{2}\right) B,\text{ \ \ \ \ }n=0,1,2,\cdots.
\label{2.1.4}
\end{equation}%
These eigenvalues are called \emph{Euclidean Landau levels}. Denote the
eigenspace of $H_{B}^{L}$ corresponding to the eigenvalue $\epsilon _{n}^{B}$
in (\ref{2.1.4}) by
\begin{equation}
\mathcal{A}_{B,n}\left( \mathbb{R}^{2}\right) =\left\{ \varphi \in
L^{2}\left( \mathbb{R}^{2},dxdy\right) ,H_{B}^{L}\left[ \varphi \right]
=\epsilon _{n}^{B}\varphi \right\} .  \label{2.1.5}
\end{equation}%
The following functions form an orthogonal basis for $\mathcal{A}%
_{B,n}\left( \mathbb{C}\right) $ \cite{HendHaimi}:
\begin{equation}
\left\{
\begin{array}{c}
e_{i,n}^{1}(z)=\sqrt{\frac{n!}{(n-i)!}}B^{\frac{i+1}{2}}z^{i}L_{n}^{\left(
i\right) }(B\left\vert z\right\vert ^{2}),\text{ \ \ }0\leq i \\
e_{j,n}^{2}(z)=\sqrt{\frac{j!}{(j+n)!}}B^{\frac{n-1}{2}}\overline{z}%
^{n}L_{j}^{\left( n\right) }(B\left\vert z\right\vert ^{2}),\text{ \ \ }%
0\leq j,%
\end{array}%
\right. \text{,}  \label{basis}
\end{equation}%
where the Laguerre polynomial is defined as
\begin{equation*}
L_{n}^{\left( \alpha \right) }(t)=\sum_{k=0}^{n}(-1)^{k}\binom{n+\alpha }{n-k%
}\frac{t^{k}}{k!}\text{, }\alpha >-1.
\end{equation*}

\begin{remark}
In his book \cite[pag. 35]{Perelomovbook}, Perelomov points out that the
basis (\ref{basis}) had been used by Feynman and Schwinger in a somewhat
different form in order to obtain an explicit expression for the matrix
elements of the displacement operator. The functions (\ref{basis}) are also
related to the complex Hermite polynomials \cite{Ismail}. They occur naturally in several
problems and different representations are used. For instance, they have
recently found applications in quantization \cite{ABG},\cite{BG}, \cite{CG},
time-frequency analysis \cite{Abreusampling}, partial differential equations
\cite{Goss} and planar point processes \cite{HendHaimi}. In the next section
we recall a characterization theorem of$\mathcal{\ }$the eigenspace $%
\mathcal{A}_{B,n}\left( \mathbb{R}^{2}\right) $ as the range of a suitable
coherent states transform of the Hilbert space $L^{2}\left( \mathbb{R}%
\right) $, originally obtained in \cite{M1}.
\end{remark}

\subsection{Coherent states for Euclidean Landau levels}

Define the Heisenberg group $\mathbb{H}$ as the Lie group whose underlying
manifold is $\mathbb{R}^{3}$ together with the group operation
\begin{equation*}
\left( x,y,r\right) \left( x\prime ,y\prime ,r\prime \right) =\left(
x+x\prime ,y+y\prime ,r+r\prime +\frac{1}{2}\left( xy\prime -x\prime
y\right) \right) \text{.}
\end{equation*}%
The continuous unitary irreducible representations of $\mathbb{H}$ are well
known \cite{Fo}. Here we consider the Schr\"{o}dinger representation $T_{B}$
of $\mathbb{H}$ on the Hilbert space $L^{2}\left( \mathbb{R},dt\right) $
\cite{Ta} defined as
\begin{equation*}
T_{B,\left( x,y,t\right) }\left[ \psi \right] \left( t\right) =\exp \left(
i\left( Bt-\sqrt{B}y\xi +\frac{B}{2}xy\right) \right) \psi \left( t-\sqrt{B}%
x\right)
\end{equation*}%
for $\left( x,y,r\right) \in \mathbb{H}$, $B>0$, $\psi \in L^{2}\left(
\mathbb{R},dt\right) $ and $t\in \mathbb{R}$. This representation is square
integrable modulo the center $\mathbb{R}$ of $\mathbb{H}$ and the Borel
section $\sigma _{0}$ of $\mathbb{H}$ over $\mathbb{H}/\mathbb{R}=\mathbb{R}%
^{2}$ which is given by $\sigma _{0}\left( x,y\right) =\left( x,y,0\right) $%
. Further, the following identity holds
\begin{equation}
\int_{\mathbb{R}^{2}}\left\langle \psi _{1},T_{B,\sigma _{0}\left(
x,y\right) }\left[ \phi _{1}\right] \right\rangle \left\langle T_{B,\sigma
_{0}\left( x,y\right) }\left[ \phi _{2}\right] ,\psi _{2}\right\rangle d\mu
\left( x,y\right) =\left\langle \psi _{1},\psi _{2}\right\rangle
\left\langle \phi _{1},\phi _{2}\right\rangle  \label{2.2.3}
\end{equation}%
for all $\psi _{1},\psi _{2},\phi _{1},\phi _{2}\in L^{2}\left( \mathbb{R}%
\right) $. Displacing the reference state
\begin{equation*}
\left\langle t\mid 0\right\rangle _{B,n}=\left( \sqrt{\pi }2^{n}n!\right) ^{-%
\frac{1}{2}}e^{-\frac{1}{2}t^{2}}H_{n}\left( t\right) \text{, }t\in \mathbb{R%
}\text{,}
\end{equation*}%
where $H_{n}\left( .\right) $ is the Hermite polynomial
\begin{equation*}
H_{n}\left( t\right) =\sum\limits_{k=0}^{\left[ n/2\right] }\frac{n!\left(
-1\right) ^{k}\left( 2t\right) ^{n-2k}}{k!\left( n-2k\right) !}\text{,}
\end{equation*}%
via the representation operator $T_{B,\sigma _{0}\left( x,y\right) }$, one
obtains a set of coherent states denoted by the kets vectors $\left\vert
\left( x,y\right) ,B,n\right\rangle $, with wave functions
\begin{equation}
\left\langle t\mid \left( x,y\right) ,B,n\right\rangle =\left( \sqrt{\pi }%
2^{n}n!\right) ^{-\frac{1}{2}}\exp \left( -i\sqrt{B}ty+i\frac{B}{2}xy-\frac{1%
}{2}\left( t-\sqrt{B}x\right) ^{2}\right) H_{n}\left( t-\sqrt{B}x\right) .
\label{2.2.5}
\end{equation}%
The following resolution of the identity
\begin{equation*}
\mathbf{1}_{L^{2}\left( \mathbb{R}\right) }=\int_{\mathbb{R}^{2}}\mid \left(
x,y\right) ,B,n\left\rangle {}\right\langle \left( x,y\right) ,B,n\mid d\mu
\left( x,y\right)
\end{equation*}%
holds as a consequence of (\ref{2.2.3}). Thus the construction of coherent
states is justified by the square integrability of representation $T_{B}$
modulo the subgroup $\mathbb{R}$ and the section $\sigma _{0}$. For $n=0$
(the lowest Euclidean Landau level), the states $\left\vert \left(
x,y\right) ,B,0\right\rangle $ coincide with the canonical coherent states
of the harmonic oscillator. The coherent states (\ref{2.2.5}) are associated
with the coherent states transform
\begin{equation*}
V_{B,n}:L^{2}\left( \mathbb{R}\right) \rightarrow L^{2}\left( \mathbb{R}%
^{2},dxdy\right)
\end{equation*}%
such that, given $\varphi \in L^{2}\left( \mathbb{R}\right) $,
\begin{equation*}
V_{B,n}\left[ \varphi \right] \left( x,y\right) :=\int_{\mathbb{R}}\overline{%
\left\langle t\mid \left( x,y\right) ,B,n\right\rangle }\varphi \left(
t\right) dt\text{.}
\end{equation*}%
Thanks to the square integrability of $T_{B}$, the transform $V_{B,n}$ is an
isometrical map. Since $V_{B,n}$ maps the Hermite functions (an orthogonal
basis of $L^{2}\left( \mathbb{R}\right) $) to the basis (\ref{basis}) (see
\cite{Abreusampling} for details) its range is exactly the eigenspace in (%
\ref{2.1.5}):
\begin{equation*}
V_{B,n}\left[ L^{2}\left( \mathbb{R}\right) \right] =\mathcal{A}_{B,n}\left(
\mathbb{R}^{2}\right) \text{.}
\end{equation*}%
Another realization of this eigenspace can be obtained by intertwining the
Landau Hamiltonian (\ref{2.1.3}) as follows
\begin{equation*}
\Delta _{B}:=e^{\frac{B}{2}z\overline{z}}\left( \frac{1}{2}H_{2B}^{L}-\frac{B%
}{2}\right) e^{-\frac{B}{2}z\overline{z}}=-\frac{\partial ^{2}}{\partial
z\partial \overline{z}}+B\overline{z}\frac{\partial }{\partial \overline{z}}.
\end{equation*}%
The space $\mathcal{A}_{B,n}\left( \mathbb{R}^{2}\right) $ then becomes
\begin{equation}
\mathcal{A}_{B,n}\left( \mathbb{C}\right) :=\left\{ \varphi \in L^{2}\left(
\mathbb{C}\text{, }e^{-Bz\overline{z}}d\mu \right) ,\Delta _{B}\varphi
=nB\varphi \right\} .  \label{2.2.9}
\end{equation}

If $B=\pi $ and $n=0$ the space (\ref{2.2.9}) is precisely the Fock-Bargmann
space of entire square integrable functions with respect to the Gaussian
measure on $\mathbb{C}$. For $n>0$, the characterization takes the form
\begin{equation*}
\widetilde{V}_{2\pi ,n}\left[ L^{2}\left( \mathbb{R}\right) \right] =%
\mathcal{A}_{\pi ,n}\left( \mathbb{C}\right)
\end{equation*}
where the coherent state transform is given explicitly by
\begin{equation*}
\widetilde{V}_{2\pi ,n}\left[ \varphi \right] \left( z\right) =e^{\frac{1}{2}%
\pi z\overline{z}}\circ V_{2\pi ,n}\left[ \varphi \right] \left( z\right)
=\left( -1\right) ^{n}B_{n}\left[ \varphi \right] \left( \sqrt{\pi }z\right)
\end{equation*}
where
\begin{equation*}
B_{n}\left[ \varphi \right] \left( w\right) =\left( -1\right) ^{n}c_{n}\int_{%
\mathbb{R}}\varphi \left( t\right) \exp \left( -\frac{1}{2}t^{2}+\sqrt{2}tw-%
\frac{1}{2}w^{2}\right) H_{n}\left( t-\frac{w+\overline{w}}{\sqrt{2}}\right)
dt.
\end{equation*}
The transform $\widetilde{V}_{2\pi ,n}$ is precisely the true polyanalytic
Bargmann transform and the space $\mathcal{A}_{\pi ,n}\left( \mathbb{C}%
\right) $ is the \textit{true-polyanalytic} space of index $n$, see \cite%
{VasiFock}, \cite{Abreusampling}, \cite{HendHaimi}.

\subsection{Completeness properties}

We want to understand the completeness properties of the coherent states
constructed in the previous section once they are labeled by a lattice $%
\Lambda \subset \mathbb{C}$. The key observation is the fact that their
completeness and basis properties are equivalent to the completeness and
basis properties of Gabor systems with Hermite functions \cite{CharlyYura}
and to sampling and uniqueness sets in true-polyanalytic spaces \cite%
{Abreusampling}. Consider the lattice
\begin{equation*}
\Lambda =\Lambda (\omega _{1},\omega _{2}):=\{m_{1}\omega _{1}+m_{2}\omega
_{2};m_{1},m_{2}\in \mathbb{Z}\}\subset \mathbb{C}
\end{equation*}%
spanned by the periods $\omega _{1}$ and $\omega _{2}\in \mathbb{C}$ with $%
\Im(\omega_{1}/\omega_{2})>0$. The size of the lattice $\Lambda $ is
the area of the parallelogram spanned by $\omega _{1}$ and $\omega _{2}$.
Identifying $\mathbb{R}^{2}$ with $\mathbb{C}$ we can write $\Lambda =\Omega
\mathbb{Z}^{2}$, where $\Omega =\left[ \omega _{1},\omega _{2}\right] $ is
an invertible $2\times 2$ matrix. The size of the lattice can now be defined
as $s(\Lambda )=\left\vert \det \Omega \right\vert $. We say that $\Lambda $
is a set of sampling for the space $\mathcal{A}_{B,n}\left( \mathbb{C}%
\right) $ if there exist constants $C_{1},C_{2}>0$ such that for all $F\in
\mathcal{A}_{B,n}\left( \mathbb{C}\right) ,$%
\begin{equation*}
C_{1}\left\Vert F\right\Vert _{\mathcal{A}_{B,n}\left( \mathbb{C}\right)
}^{2}\leq \sum\limits_{\lambda \in \Lambda }\left\vert F\left( \lambda
\right) \right\vert ^{2}e^{-B\left\vert \lambda \right\vert ^{2}}\leq
C_{2}\left\Vert F\right\Vert _{\mathcal{A}_{B,n}\left( \mathbb{C}\right)
}^{2}\text{.}
\end{equation*}%
Given a point $\left( q,p\right) $ in the phase space $\mathbb{R}^{2}$, the
corresponding time-frequency shift is
\begin{equation*}
\pi _{\left( q,p\right) }\left[ f\right] \left( t\right) =e^{2\pi
ipt}f\left( t-q\right) ,t\in \mathbb{R}\text{.}
\end{equation*}%
Let $h_{n}\left( t\right) $ denote a Hermite function. The set $G\left(
h_{n},\Lambda \right) :=\left\{ \pi _{\left( q,p\right) }h_{n},\left(
q,p\right) \in \mathbb{R}\right\} $ is a Gabor frame or a Weyl-Heisenberg
frame in $L^{2}\left( \mathbb{R}\right) $ whenever there exist constants $%
C_{1},C_{2}>0$ such that
\begin{equation*}
C_{1}\left\Vert f\right\Vert _{L^{2}\left( \mathbb{R}\right) }^{2}\leq
\sum\limits_{\left( q,p\right) \in \Lambda }\left\vert \left\langle f,\pi
_{\left( q,p\right) }\left[ h_{n}\right] \right\rangle _{L^{2}\left( \mathbb{%
R}\right) }\right\vert ^{2}\leq C_{2}\left\Vert f\right\Vert _{L^{2}\left(
\mathbb{R}\right) }^{2}.
\end{equation*}%
\ It follows from the lower inequality that if $G\left( h_{n},\Lambda
\right) $\ is a frame then $G\left( h_{n},\Lambda \right) $\ is complete.
For simplicity, we consider the square lattice $\Lambda _{\omega }:=\omega
\left( \mathbb{Z+}i\mathbb{Z}\right) $, $\omega \in \mathbb{R}$. In this
case $s(\Lambda _{\omega })=\omega ^{2}$. For $B=\pi $, it was proved that
the lattice $\Lambda _{\omega }$ is a set of sampling for $\mathcal{A}_{\pi
,n}\left( \mathbb{C}\right) $ if and only if $G\left( h_{n},\Lambda _{\omega
}\right) $ is a Gabor frame, see \cite{Abreusampling}. The following result
is a consequence of combining this identification with relatively recent
results from time-frequency analysis.

\begin{theorem}
\textit{Let }$(\left\vert \left( x,y\right) ,\pi ,n\right\rangle )_{\left(
x,y\right) \in \mathbb{R}^{2}}$\textit{\ be a system of coherent states
attached to the }$nth$\textit{\ Landau level defined in (\ref{2.2.5}). Then,
the following holds:}
\end{theorem}

$\left( i\right) $ \textit{If }$\omega ^{2}<\frac{1}{n+1}$\textit{\ then the
system }$(\left\vert \left( x,y\right) ,\pi ,n\right\rangle )_{\left(
x,y\right) \in \Lambda _{\omega }}$\textit{\ is complete.}

$\left( ii\right) $\textit{\ If }$\omega ^{2}>1$\textit{\ then the system }$%
(\left\vert \left( x,y\right) ,\pi ,n\right\rangle )_{\left( x,y\right) \in
\Lambda _{\omega }}$ \textit{is} \textit{incomplete.}

\textbf{Proof. }The completeness property (i) follows from the fact that if%
\textit{\ }$\omega ^{2}<\frac{1}{n+1}$, then $G\left( h_{n},\Lambda \right) $
is a Gabor frame \cite{CharlyYura}, therefore complete. The property (ii) is
a consequence of the fact that, if $\omega ^{2}>1$, then a Gabor system
cannot be complete \cite{RamStee}.

\begin{remark}
In the case $n=0$ it is a classical result \cite{Perelomov}, \cite%
{BargmannEt} that the systems are complete if $\omega ^{2}\leq 1$ and
incomplete if $\omega ^{2}>1$. The above result is an extension of these
results to coherent states attached to higher Euclidean Landau levels $%
\epsilon _{n}^{\pi },$ $n=1,2,3,...$ .
\end{remark}

\begin{remark}
For $n>0$ there is still a considerable gap between conditions (i) and (ii).
Finding a whole description of the completeness and frame properties of
Gabor systems indexed by lattices is a highly non-trivial problem which has
been subject of study since \cite{CharlyYura}. The very recent preprint \cite%
{completeness} seems to answer the question in the case of rational lattices.
\end{remark}

\begin{remark}
The Landau Hamiltonian arises in the two-dimensional quantized Hall effect.
A coherent states formalism for the study of this problem has been developed
by projecting the higher order states in the lowest Landau Level, which can
be modelled by analytic functions \cite{GirvJach}. It is reasonable to
expect that the discrete coherent states associated with higher Landau
Levels may provide an alternative formalism.
\end{remark}

\section{Hyperbolic Landau levels}

\subsection{Hyperbolic Landau levels}

In the hyperbolic setting, the configuration space is now the Poincar\'{e}
upper half-plane $\mathbb{C}^{+}=\left\{ z\in \mathbb{C},\Im%
z>0\right\} $. It is a complete two-dimensional simply connected Riemannian
manifold of constant negative curvature $R=-1$, endowed with the metric $%
ds^{2}=y^{-2}\left( dx^{2}+dy^{2}\right) $, where $z=x+iy$. A constant
homogeneous magnetic field on $\mathbb{C}^{+}$ is given by a $2$-form $d\mu
_{B}$ defined as
\begin{equation*}
d\mu _{B}=\frac{2B}{y^{2}}dxdy
\end{equation*}%
where $B$ is the field intensity. The form $d\mu _{B}$ is exact and any $1$%
-form $A$ such that $d\mu _{B}=dA$ is called a vector potential related to $%
d\mu _{B}$. For our purposes it is convenient to choose $A=2By^{-1}dx.$ In
suitable units and up to an additive constant, the Schr\"{o}dinger operator
describing the dynamics of a charged particle moving on $\mathbb{C}^{+}$
under the action of the magnetic field $B$ is given by \cite{Comtet}
\begin{equation*}
H_{B}:=y^{2}\left( \frac{\partial ^{2}}{\partial x^{2}}+\frac{\partial ^{2}}{%
\partial y^{2}}\right) -2iBy\frac{\partial }{\partial x}.
\end{equation*}%
Different aspects of the spectral analysis of the operator $H_{B}$ have been
studied by many authors, (see \cite{Groshe}, \cite{Comtet} or, for a more
mathematical approach, \cite{Pa}). We list here the following important
properties.

$\left( i\right) $ $H_{B}$ is an elliptic densely defined operator on the
Hilbert space $L^{2}\left( \mathbb{C}^{+},d\mu _{B}\right) $, with a unique
self-adjoint realization that we denote also by $H_{B}$.

$\left( ii\right) $ The spectrum\ of $H_{B}$ in $L^{2}\left( \mathbb{C}%
^{+},d\mu _{B}\right) $ consists of two parts:\textit{\ }a continuous part $%
\left[ 1/4,+\infty \right[ $, corresponding to \textit{scattering states}
and a finite number of eigenvalues with infinite degeneracy (\textit{%
hyperbolic Landau levels}) of the form
\begin{equation}
\epsilon _{n}^{B}:=(B-n)\left( 1-B+n\right) ,n=0,1,2,\cdots ,\lfloor B-\frac{%
1}{2}\rfloor \text{.}  \label{3.1.2}
\end{equation}%
The finite part of the spectrum exists provided $2B>1$. The notation $%
\lfloor a\rfloor $ stands for the greatest integer not exceeding $a.$

$\left( iii\right) $ For each fixed eigenvalue $\epsilon _{n}^{B}$, we
denote by
\begin{equation}
\mathcal{E}_{n}^{B}\left( \mathbb{C}^{+}\right) =\left\{ \Phi \in
L^{2}\left( \mathbb{C}^{+},d\mu _{B}\right) ,H_{B}\Phi =\epsilon
_{n}^{B}\Phi \right\}  \label{3.1.3}
\end{equation}%
the corresponding eigenspace. Its reproducing kernel is given by
\begin{equation*}
K_{n,B}\left( z,\zeta \right) =\frac{\left( -1\right) ^{n}\Gamma \left(
2B-n\right) }{n!\Gamma \left( 2B-2n\right) }\left( \frac{\left\vert z-%
\overline{\zeta }\right\vert ^{2}}{4\Im z\Im\zeta }\right)
^{-B+m}\left( \frac{\zeta -\overline{z}}{z-\overline{\zeta }}\right) ^{B}
\end{equation*}%
\begin{equation*}
\times _{2}F_{1}\left( -2B-m,-m,2B-2m,\frac{4\Im z\Im\zeta }{%
\left\vert z-\overline{\zeta }\right\vert ^{2}}\right)
\end{equation*}%
where $_{2}F_{1}$ is the Gauss hypergeometric function.

\begin{remark}
The condition $2B>1$ ensuring the existence of the discrete eigenvalues
means that the magnetic field has to be strong enough to capture the
particle in a closed orbit. If this condition is not fulfilled the motion
will be unbounded and the particle will escape to infinity. More precisely,
the orbit of the particle will intercept the upper half-plane boundary whose
points stand for points at infinity\ \cite[pg. 189]{Comtet}.\ To the
eigenvalues in (\ref{3.1.2}) below the continuous spectrum correspond
eigenfunctions which are called \textit{bound states. This terminology is
due to the fact that} the particle in such a bound state cannot leave the
system without additional energy.
\end{remark}

\subsection{Bergman spaces}

For $n=0$, the reproducing kernel of $\mathcal{E}_{0}^{B}\left( \mathbb{C}%
^{+}\right) $ reduces to
\begin{equation*}
K_{0,B}\left( z,\zeta \right) =e^{i\pi B}4^{B}\frac{\left( \Im z\Im%
\zeta \right) ^{B}}{\left( z-\overline{\zeta }\right) ^{2B}}\text{.}
\end{equation*}%
This is the reproducing kernel for the $(2B-2)$-weighted Bergman space $%
A_{2B-1}\left( \mathbb{C}^{+}\right) $, constituted by analytic functions $f$
on the upper half-plane with finite norm%
\begin{equation*}
\left\Vert f\right\Vert _{A_{2B-1}\left( \mathbb{C}^{+}\right)
}=\int\limits_{\mathbb{C}^{+}}\left\vert f\left( z\right) \right\vert
^{2}y^{2B-2}dxdy<+\infty.
\end{equation*}%
Thus, $\mathcal{E}_{0}^{B}\left( \mathbb{C}^{+}\right) $ coincides with $%
A_{2B-2}\left( \mathbb{C}^{+}\right) $.

An important fact to be used in the appendix proof of the main results is
the following. Note also that for a general weight $\nu $, the Bergman space
$A_{\nu }\left( \mathbb{C}^{+}\right) $ is  connected to the space $%
L^{2}\left( \mathbb{R}^{+},t^{-1}dt\right) $\ by the integral transform
defined as
\begin{equation}
Ber_{\nu }\left[ h\right] \left( z\right) =\int\nolimits_{0}^{+\infty }t^{%
\frac{\nu +3}{2}}h\left( t\right) e^{izt}dt  \label{3.1.6}
\end{equation}%
see, for instance \cite{Dau}, \cite{WavFram}. This provides an isometric
isomorphism
\begin{equation*}
Ber_{\nu }:L^{2}\left( \mathbb{R}^{+},t^{-1}dt\right) \rightarrow A_{\nu
}\left( \mathbb{C}^{+}\right) \text{.}
\end{equation*}%
The transform is onto because one can deduce from the special function
formula%
\begin{equation}
\int_{0}^{\infty }t^{\alpha }L_{m}^{\alpha }(t)e^{-tu}dt=\frac{\Gamma
(m+1+\alpha )}{m!}\left( \frac{u-1}{u}\right) ^{m}\frac{1}{u^{\alpha +1}}
\label{intLaguerre}
\end{equation}%
that the Laguerre functions are mapped to a basis of $A_{\nu }\left( \mathbb{%
C}^{+}\right) $ formed by rational functions which are further mapped to the
unit disc by a conformal map. Some details and remarks about these
calculations are given in \cite{AbreuRemarks} and \cite{DGM}.

\subsection{The affine group acting on the Poincar\'{e} half-plane}

For our purposes we will recall a characterization theorem of $\mathcal{E}%
_{n}^{B}\left( \mathbb{C}^{+}\right) $ as the range under a suitable
coherent states transform $W_{B,n}$ defined on the Hilbert space $\mathcal{H}%
:=$ $L^{2}\left( \mathbb{R}^{+},t^{-1}dt\right) $. We start with the
identification of the Poincar\'{e} upper half-plane $\mathbb{C}^{+}$ with
the affine group $\mathbf{G}=\mathbb{R}\times \mathbb{R}^{+}$, by setting $%
z=x+iy\equiv \left( x,y\right) $. The group law of $\mathbb{G}$ is $\left(
x,y\right) .\left( x\prime ,y\prime \right) =\left( x+yx\prime ,yy\prime
\right) $. $\mathbf{G}$ is a locally compact unimodular group with the left
Haar measure $d\mu \left( x,y\right) =y^{-2}dxdy$ and modular function $%
\Delta \left( x,y\right) =y^{-1}$. By this identification the space $%
L^{2}\left( \mathbf{G},d\mu \right) $ coincides with the space $L^{2}\left(
\mathbb{C}^{+},d\mu _{B}\right) $. We shall consider one of the two
inequivalent infinite dimensional unitary irreducible representations of the
affine group $\mathbf{G}$, denoted $\pi _{+}$, realized on the Hilbert space
$\mathcal{H}$ as
\begin{equation*}
\pi _{+}\left( x,y\right) \left[ \varphi \right] \left( t\right) :=\exp
\left( ixt/2\right) \varphi \left( yt\right) ,\text{ \ }\varphi \in \mathcal{%
H},\text{ \ }t\in \mathbb{R}^{+}\text{.}
\end{equation*}%
This representation is square integrable since it is easy to find a vector $%
\phi _{0}\in \mathcal{H}$ such that the function $\left( x,y\right) \mapsto
\left\langle \pi _{+}\left( x,y\right) \left[ \phi _{0}\right] ,\phi
_{0}\right\rangle _{\mathcal{H}}$ belongs to $L^{2}\left( \mathbf{G},d\mu
\right) $. This condition can also be expressed by saying that the
self-adjoint operator $K:\mathcal{H\rightarrow H}$ defined as $K\left[ \psi %
\right] (.)=(.)^{-\frac{1}{2}}\psi \left( .\right) $ gives
\begin{equation*}
\int\limits_{\mathbf{G}}d\mu \left( x,y\right) \left\langle \varphi
_{1},\pi _{+}\left( x,y\right) \left[ \psi _{1}\right] \right\rangle
\left\langle \pi _{+}\left( x,y\right) \psi _{2},\left[ \varphi _{2}\right]
\right\rangle =\left\langle \varphi _{1},\varphi _{2}\right\rangle
\left\langle K^{\frac{1}{2}}\left[ \psi _{1}\right] ,K^{\frac{1}{2}}\left[
\psi _{2}\right] \right\rangle \text{,}
\end{equation*}%
for all $\psi _{1},\psi _{2},\varphi _{1},\varphi _{2}\in \mathcal{H}$. The
operator $K$ is unbounded because $\mathbf{G}$ is not unimodular \cite{DM}.
We will also use the notation
\begin{equation*}
\pi _{+}^{1}\left( x,y\right) \left[ \varphi \right] \left( t\right) :=y^{%
\frac{1}{2}}\exp \left( ixt/2\right) \varphi \left( yt\right) ,\text{ \ }%
\varphi \in \mathcal{H},\text{ \ }t\in \mathbb{R}^{+}
\end{equation*}%
such that
\begin{equation}
\pi _{+}\left( x,y\right) \left[ (.)^{\frac{1}{2}}\varphi \left( .\right) %
\right] (t)=t^{\frac{1}{2}}\pi _{+}^{1}\left( x,y\right) \left[ \varphi %
\right] \left( t\right)  \label{1+}
\end{equation}%
and also, for functions $\Phi $ such that their Fourier transform belongs to
$L^{2}\left( \mathbb{R}^{+}\right) $ (this is essentially the Hardy space
where the wavelet transformation is often defined),%
\begin{equation*}
\pi ^{wav}\left( x,y\right) \left[ \Phi \right] \left( t\right) =y^{-\frac{1%
}{2}}\Phi \left( y^{-1}\left( t-x\right) \right) \text{.}
\end{equation*}%
For shortness of notations, in some situations we will represent the point $%
(x,y)$ by the complex number $z=x+iy$, often with no explicit mention.

\subsection{Coherent states for higher hyperbolic Landau levels}

Now, as in \cite{Mouayn}, we consider a set of coherent states denoted by
the ket vectors $\left\vert \left( x,y\right) ,B,n\right\rangle $ and
obtained by displacing, via the representation operator $\pi _{+}\left(
x,y\right) $, the reference state vector $\left\vert 0\right\rangle _{B,n}$
in the Hilbert space $\mathcal{H}$ with wave function given by
\begin{equation*}
\left\langle t\mid 0\right\rangle _{B,n}=\left( \frac{\Gamma \left(
2B-n\right) }{n!}\right) ^{-\frac{1}{2}}t^{B-n}e^{-\frac{1}{2}%
t}L_{n}^{\left( 2B-2n-1\right) }\left( t\right) \text{.}
\end{equation*}%
Precisely,
\begin{equation}
\left\vert \left( x,y\right) ,B,n\right\rangle :=\pi _{+}\left(
x,y\right)\left|0\right\rangle _{B,n}\text{.}  \label{3.2.4}
\end{equation}%
The wave functions of the coherent states $(\ref{3.2.4})$\ are given by
\begin{equation}
\left\langle t\mid \left( x,y\right) ,B,n\right\rangle =\left( \frac{\Gamma
\left( 2B-n\right) }{n!}\right) ^{-\frac{1}{2}}\left( ty\right) ^{B-n}e^{-%
\frac{1}{2}t\left( y-ix\right) }L_{n}^{\left( 2B-2n-1\right) }\left(
ty\right) \text{.}  \label{3.2.5}
\end{equation}%
These coherent states are completely justified by the square integrability
of the unitary irreducible representation $\pi _{+}$ and if follows from the
special function formula (\ref{intLaguerre}) that we have a resolution of
the identity for the space $\mathcal{H=}$ $L^{2}\left( \mathbb{R}%
^{+},t^{-1}dt\right) $:
\begin{equation*}
\mathbf{1}_{\mathcal{H}}=c_{B,n}^{-1}\int\limits_{\mathbf{G}}d\mu \left(
x,y\right) \left\vert \left( x,y\right) ,B,n\right\rangle \left\langle
\left( x,y\right) ,B,n\right\vert \text{,}
\end{equation*}%
where $c_{B,n}=\left( 2\left( B-n\right) -1\right) ^{-1}$. The coherent
states (\ref{3.2.4}) are associated with the coherent states transform
\begin{equation}
W_{B,n}\left[ \varphi \right] \left( x,y\right) =c_{B,n}^{-\frac{1}{2}%
}\int\limits_{\mathbb{R}^{+}}\overline{\left\langle t\mid \left( x,y\right)
,B,n\right\rangle }\varphi \left( t\right) \frac{dt}{t}.  \label{22}
\end{equation}%
The range of the map $W_{B,n}:L^{2}\left( \mathbb{R}^{+},t^{-1}dt\right)
\rightarrow L^{2}\left( \mathbb{C}^{+},d\mu _{B}\right) $ is the eigenspace (%
\ref{3.1.3}):
\begin{equation*}
W_{B,n}\left[ L^{2}\left( \mathbb{R}^{+},t^{-1}dt\right) \right] =\mathcal{E}%
_{n}^{B}\left( \mathbb{C}^{+}\right)
\end{equation*}%
for every $n\in \mathbb{Z}_{+}\cap \left[ 0,B-\frac{1}{2}\right] $ provided
that $2B>1$.

\begin{remark}
Note that, for $n=0$, the lowest hyperbolic Landau level, the states $%
\left\vert \left( x,y\right) ,B,0\right\rangle $ coincide with the well
known \textit{affine} coherent states \cite{KL}.
\end{remark}

\subsection{Wavelet transforms with Laguerre functions}

In this subsection we write the coherent states of the previous section in
terms of wavelet transforms with analyzing wavelets $\Phi _{n}^{\alpha }$
defined via the Fourier transforms in terms of Laguerre polynomials $%
L_{n}^{\alpha }$\ as%
\begin{equation}
\mathcal{F}\Phi _{n}^{\alpha }(t)=t^{\frac{\alpha +1}{2}}e^{-t}L_{n}^{\alpha
}(2t)\text{.}  \label{laguerre}
\end{equation}%
Some of the structural properties of $\Phi _{n}^{\alpha }$ that will be key
in our approach are a consequence of their explicit formula, which displays $%
\Phi _{n}^{\alpha }$ as linear combinations of $\{\Phi _{n}^{\alpha
+2k}\}_{k=0}^{n}$:
\begin{equation*}
\Phi _{n}^{\alpha }(t)=\sum_{k=0}^{n}\frac{(-2)^{k}}{k!}\left(
\begin{array}{c}
n+\alpha \\
n-k%
\end{array}%
\right) \Phi _{0}^{\alpha +2k}(t)\text{.}
\end{equation*}

Now, let\textbf{\ }$\varphi \in L^{2}\left( \mathbb{R}^{+},t^{-1}dt\right) $%
. Combining (\ref{22}) and (\ref{3.2.5}) gives
\begin{equation*}
W_{B,n}\left[ \varphi \right] \left( x,y\right) =c_{B,n}^{-\frac{1}{2}%
}\left( \frac{\Gamma \left( 2B-n\right) }{n!}\right) ^{-\frac{1}{2}%
}\int\limits_{\mathbb{R}^{+}}\left( ty\right) ^{B-n}e^{-\frac{1}{2}t\left(
y+ix\right) }L_{n}^{\left( 2B-2n-1\right) }\left( ty\right) \varphi \left(
t\right) \frac{dt}{t}\text{.}
\end{equation*}%
With $z=x+iy,$ we have that $-\frac{1}{2}t\left( y+ix\right) =\overline{%
\frac{1}{2}\xi iz}$. Set $\gamma _{B,n}=c_{B,n}\left( n!\right) ^{-1}\Gamma
\left( 2B-n\right) $ and rewrite the above as
\begin{equation}
W_{B,n}\left[ \varphi \right] \left( x,y\right) =\gamma _{B,n}^{-\frac{1}{2}%
}\int\limits_{\mathbb{R}^{+}}\varphi \left( t\right) \overline{\left( \left(
ty\right) ^{B-n}e^{\frac{1}{2}\xi iz}L_{n}^{\left( 2B-2n-1\right) }\left(
ty\right) \right) }\frac{dt}{t}  \label{3.3.13}
\end{equation}%
Since $\pi _{+}\left( x,y\right) \left[ \left( .\right) ^{\frac{1}{2}%
}l_{n}^{2B-2n-1}(.)\right] \left( t\right) =\left( ty\right) ^{B-n}e^{\frac{1%
}{2}tiz}L_{n}^{\left( 2B-2n-1\right) }\left( ty\right) $, then (\ref{3.3.13}%
) becomes
\begin{eqnarray}
W_{B,n}\left[ \varphi \right] \left( x,y\right) &=&\gamma _{B,n}^{-\frac{1}{2%
}}\int\limits_{\mathbb{R}^{+}}\varphi \left( t\right) \overline{\left( \pi
_{+}\left( x,y\right) \left[ \left( .\right) ^{\frac{1}{2}}l_{n}^{2B-2n-1}(.)%
\right] \right) }\left( t\right) \frac{dt}{t}  \notag \\
&=&\gamma _{B,n}^{-\frac{1}{2}}\left\langle \varphi ,\pi _{+}\left(
x,y\right) \left[ \left( .\right) ^{\frac{1}{2}}l_{n}^{2B-2n-1}(.)\right]
\right\rangle _{L^{2}\left( \mathbb{R}^{+},\frac{dt}{t}\right) }\text{.}
\label{3316}
\end{eqnarray}%
Since $\pi _{+}\left( x,y\right) \left[ \left( .\right) ^{\frac{1}{2}}\phi
\left( .\right) \right] \left( t\right) =t^{\frac{1}{2}}\pi _{+}^{1}\left(
x,y\right) \left[ \phi \left( .\right) \right] \left( t\right) $, then (\ref%
{3316}) becomes
\begin{equation*}
W_{B,n}\left[ \varphi \right] \left( x,y\right) =\gamma _{B,n}^{-\frac{1}{2}%
}\left\langle (.)^{-\frac{1}{2}}\varphi (.),\pi _{+}^{1}\left( z\right) %
\left[ l_{n}^{2B-2n-1}\right] (.)\right\rangle _{L^{2}\left( \mathbb{R}%
^{+},dt\right) }\text{.}
\end{equation*}%
If $\varphi \in L^{2}\left( \mathbb{R}^{+},t^{-1}dt\right) $, then $\mathcal{%
F}^{-1}\left( t^{-\frac{1}{2}}\varphi \right) \ $is in $H^{2}\left( \mathbb{C%
}^{+}\right) $ and the scalar product above may also be written as
\begin{equation*}
\left\langle (.)^{-\frac{1}{2}}\varphi (.),\pi _{1}^{+}\left( z\right) \left[
l_{n}^{2B-2n-1}\right] (.)\right\rangle _{L^{2}\left( \mathbb{R}%
^{+},dt\right) }=\mathcal{W}_{\Phi _{n}^{2\left( B-n\right) -1}}\left[
\mathcal{F}^{-1}\left( \left( .\right) ^{-\frac{1}{2}}\varphi \right) \right]
\left( z\right) \text{,}
\end{equation*}%
where $\mathcal{W}_{\Phi _{n}^{2\left( B-n\right) -1}}$ stands for the \emph{%
wavelet transformation} \cite{Dau}, defined as
\begin{equation*}
\mathcal{W}_{\Phi }\left[ \varphi \right] \left( x,y\right) =\left\langle
\varphi ,\pi _{z}\Phi \right\rangle _{L^{2}\left( \mathbb{R}\right) }\text{,
\ \ \ }z=x+iy\text{, \ \ }y>0\text{,}
\end{equation*}%
where $\mathcal{F}\Phi \in L^{2}\left( \mathbb{R}^{+},t^{-1}dt\right) $. The
two transforms are related as follows
\begin{equation}
W_{B,n}\left[ \varphi \right] \left( x,y\right) =\gamma _{B,n}^{-\frac{1}{2}}%
\mathcal{W}_{\Phi _{n}^{2\left( B-n\right) -1}}\left[ \mathcal{F}^{-1}\left(
\left( .\right) ^{-\frac{1}{2}}\varphi (.)\right) \right] \left( x,y\right)
\text{.}  \label{3.3.22}
\end{equation}

\begin{remark}
This also means that we have another realization of the bound states space $%
\mathcal{E}_{n}^{B}\left( \mathbb{C}^{+}\right) $ in (\ref{3.1.3}) as the
image of the Hardy space $H\left( \mathbb{C}^{+}\right) $ under the wavelet
transform $\mathcal{W}_{\Phi _{n}^{2\left( B-n\right) -1}}$.
\end{remark}

With the help of the transform $Ber_{\nu }$ in (\ref{3.1.6}), we will be
able to express the transform $W_{B,n}\left[ f\right] $ of any function $f$
in $L^{2}\left( \mathbb{R}^{+},t^{-1}dt\right) $ as a combination of
derivatives of an analytic function.

\begin{proposition}
If\textit{\ }$f\in L^{2}\left( \mathbb{R}^{+},t^{-1}dt\right) $\textit{, then%
}
\begin{equation*}
W_{B,n}\left[ f\right] \left( z\right) =\gamma _{B,n}^{-\frac{1}{2}%
}\sum\limits_{k=0}^{n}\frac{\left( 2i\right) ^{k}}{k!}\left(
\begin{array}{c}
2B-n-1 \\
n-k%
\end{array}%
\right) y^{B-n-\frac{1}{2}+k}F^{\left( k\right) }\left( z\right) \text{,}
\end{equation*}%
\textit{where }$F\left( z\right) =Ber_{2\left( B-n\right) -1}\left[ f\right]
(z)$\textit{\ belongs to the weighted Bergman space }$A_{2\left( B-n\right)
-1}\left( \mathbb{C}^{+}\right) $\textit{.}
\end{proposition}

\textbf{Proof.} Take $f\in L^{2}\left( \mathbb{R}^{+},t^{-1}dt\right) $. Then the function $%
u=\mathcal{F}^{-1}\left( t^{-\frac{1}{2}}f\right) \in H^{2}\left( \mathbb{C}%
^{+}\right) $. Write $F:=Ber_{\nu }\left[ f\right] $, where $\nu =2\left(
B-n\right) -1$. In \cite[pg. 256]{WavFram}, it is shown that the wavelet
transform of $u$ decomposes in terms of derivatives of the analytic function
$F\in A_{2\left( B-n\right) -1}\left( \mathbb{C}^{+}\right) $ as
\begin{equation}
\widetilde{\mathcal{W}}_{\Phi _{n}^{2\left( B-n\right) -1}}u\left( z\right)
=\sum\limits_{k=0}^{n}\frac{\left( 2i\right) ^{k}}{k!}\left(
\begin{array}{c}
2B-n-1 \\
n-k%
\end{array}%
\right) y^{B-n-\frac{1}{2}+k}F^{\left( k\right) }\left( z\right) \text{.}
\label{3.3.23}
\end{equation}%
Recalling the relation (\ref{3.3.22}) between the two transforms, we may
rewrite (\ref{3.3.23}) as
\begin{equation}
W_{B,n}f\left( x,y\right) =\gamma _{B,n}^{-\frac{1}{2}}\sum\limits_{k=0}^{n}%
\frac{\left( 2i\right) ^{k}}{k!}\left(
\begin{array}{c}
2B-n-1 \\
n-k%
\end{array}%
\right) y^{B-n-\frac{1}{2}+k}F^{\left( k\right) }\left( z\right) \text{.}
\tag{3.3.24}
\end{equation}%
This completes the proof.
\subsection{Fuchsian groups and their automorphic forms}

Let $I_{2}$ be the identity matrix. Since one can identify the Poincar\'{e}
half-plane $\mathbb{C}^{+}$\ with the quotient group
\begin{equation*}
PSL\left( 2,\mathbb{R}\right) :=SL\left( 2,\mathbb{R}\right) /\left\{ \pm
I_{2}\right\} \text{,}
\end{equation*}%
also known as the group of M\"{o}bius transformations, the subgroups of $%
PSL\left( 2,\mathbb{R}\right) $, known as \emph{Fuchsian groups}, describe
the isometries of the hyperbolic metric of $\mathbb{C}^{+}$. Since the
nontrivial dynamics of a particle in the upper half-plane is induced by its
tesselation by discrete subgroups, we want to understand the completeness
properties of the coherent states introduced in the previous section, once
they are labelled by Fuchsian groups. Thus, we need to recall some basic
facts about Fuchsian groups and their associated automorphic forms. Consider
the group $SL\left( 2,\mathbb{R}\right) $ of real $2\times 2$ matrices with
determinant one, acting on $\mathbb{C}^{+}$ according to the rule
\begin{equation*}
g.z=\frac{az+b}{cz+d},\text{ \ \ \ \ \ \ \ \ \ }g=\left(
\begin{array}{cc}
a & b \\
c & d%
\end{array}%
\right) \in SL\left( 2,\mathbb{R}\right) \text{.}
\end{equation*}%
Notice that $g$ and $-g$ have the same action on $\mathbb{C}^{+}$.

A \emph{Fuchsian group }$G$ is a discrete subgroup of $PSL\left( 2,\mathbb{R}%
\right) $. The most important example is the modular group $PSL\left( 2,%
\mathbb{Z}\right) =SL\left( 2,\mathbb{Z}\right) /\left\{ \pm I_{2}\right\} $%
, where%
\begin{equation*}
SL\left( 2,\mathbb{Z}\right) =\left\{ \left(
\begin{array}{cc}
a & b \\
c & d%
\end{array}%
\right) :a,b,c,d\in \mathbb{Z}\text{, }ad-bc=1\right\} \text{.}
\end{equation*}%
An important class is provided by the congruence groups of order $n$, $G(n)$%
,
\begin{equation*}
G\left( n\right) =\left\{ \left(
\begin{array}{cc}
a & b \\
c & d%
\end{array}%
\right) \in SL\left( 2,\mathbb{Z}\right) :\left(
\begin{array}{cc}
a & b \\
c & d%
\end{array}%
\right) =\pm I(\mod n)\right\} \text{.}
\end{equation*}%
Further terminology will be required. The $G$-orbit $Gz$ of a point $z\in
\mathbb{C}^{+}$ under the action of the group $G$ is%
\begin{equation*}
Gz=\{gz:g\in G\}\text{.}
\end{equation*}%
A \textit{fundamental domain} for a Fuchsian group $G$ is a closed set $D$ $%
\subset $ $\mathbb{C}^{+}$ such that $D$ is the closure of its interior $%
D^{0}$, no two points of $D^{0}$ lie in the same $G$-orbit and the images of
$D$ under $G$ cover $\mathbb{C}^{+}$. For instance, a fundamental domain for
$G=PSL\left( 2,\mathbb{Z}\right) $ is given by%
\begin{equation*}
D=\left\{ z\in \mathbb{C}^{+}:\left\vert z\right\vert \geq 1\text{ and }%
\left\vert \Re z\right\vert \leq \frac{1}{2}\right\} \text{.}
\end{equation*}%
In the hyperbolic model, the orbit of an element $z\in D$\ will replace the
role of the lattice $\Lambda \left( \omega _{1},\omega _{2}\right) $\ in the
Euclidean model of the previous section, while the fundamental domain $D$
replaces the role of the parallelogram spanned by $\omega _{1}\ $and $\omega
_{2}$. We will restrict to Fuchsian groups such that $D$\ has finite
hyperbolic area. In this case, $D$ can be chosen as a polygon with \emph{an
even number }$2k$\emph{\ of sides}. The sides, grouped in pairs, are
equivalent with respect to the action of $G$. The vertices of the polygon
are joined in cycles of vertices which are equivalent to each other. If the
region is a polygon with vertices lying on the boundary of $\mathbb{C}^{+}$,
the cycle is called \emph{parabolic }(often referred to in the literature as
\emph{cusps}), otherwise it is called \emph{elliptic}. Let $r$ be the total
number of cycles and $e_{1},....,e_{r}$ be the orders of the inequivalent
elliptic points of $G$. Joining equivalent vertices and cycles, leads to the
construction of the Riemann surface\ $G\setminus \mathbb{C}^{+}$, whose
\emph{genus }$\mathcal{G}$ is given by $2\mathcal{G}=1+k-r$. The set $(%
\mathcal{G},r,e_{1},....,e_{r})$ is called the signature of the group $G$.
It contains information to compute the area $S_{G}$ of the fundamental
domain $D$:%
\begin{equation}
S_{G}=2\pi \left[ 2\mathcal{G}-2+\sum_{l=1}^{r}\lfloor 1-\frac{1}{e_{l}}%
\rfloor \right] \text{.}  \label{area}
\end{equation}%
Now we introduce the notion of an automorphic form associated with $G$.\ For
all $m\in \mathbb{Z}$, $z\in $ $\mathbb{C}^{+}$ and any function $f$ with
domain $\mathbb{C}^{+}$, let
\begin{equation*}
\left( f\mid _{m}g\right) \left( z\right) =\left( cz+d\right) ^{-2m}f\left(
g.z\right) \text{, \ }g=\left(
\begin{array}{cc}
a & b \\
c & d%
\end{array}%
\right) \in SL\left( 2,\mathbb{R}\right) \text{.}
\end{equation*}%
An \textit{automorphic form} of weight $m$ with respect to a Fuchsian group $%
G$ is a meromorphic function $f$ on $\mathbb{C}^{+}$ such that
\begin{equation*}
\left( f\mid _{m}g\right) =f\text{,}
\end{equation*}%
for all $g\in G$. The number $N$ of zeros of $f$ inside the fundamental
domain $D$ of the group $G$\ is given by Poincar\'{e}'s formula%
\begin{equation}
N=m\frac{S_{G}}{2\pi }\text{.}  \label{Poincare}
\end{equation}%
The set of all automorphic forms of weight $m$ is denoted by $\Omega
_{G}^{m}\left( \mathbb{C}^{+}\right) $. Consider also $\mathfrak{H}%
ol_{G}^{m}\left( \mathbb{C}^{+}\right) $, the set of functions $f\in \Omega
_{G}^{m}\left( \mathbb{C}^{+}\right) $ holomorphic on $\mathbb{C}^{+}$
(including all cusps of $G$). We write $\mathfrak{C}_{G}^{m}\left( \mathbb{C}%
^{+}\right) $ for the set of functions $f\in \Omega _{G}^{m}\left( \mathbb{C}%
^{+}\right) $ which are zero at all cusps of $G$ (the so-called cusp forms).
The inclusions among these spaces are the following:
\begin{equation*}
\mathfrak{C}_{G}^{m}\left( \mathbb{C}^{+}\right) \subset \mathfrak{H}%
ol_{G}^{m}\left( \mathbb{C}^{+}\right) \subset \Omega _{G}^{m}\left( \mathbb{%
C}^{+}\right) .
\end{equation*}%
The dimension $\dim \mathfrak{H}ol_{G}^{m}\left( \mathbb{C}^{+}\right) $ is
known explicitly \cite[p. 46 , Theorem 2. 23]{Shimura} in terms of $m$, the
genus $\mathcal{G}$ of the Riemann surface\ $G\setminus \mathbb{C}^{+}$, the
orders of the inequivalent elliptic points of $G$. Assuming that all cusps
of $G$ are equivalent,
\begin{equation}
\dim \mathfrak{H}ol_{G}^{m}\left( \mathbb{C}^{+}\right) =\left\{
\begin{array}{cc}
\left( 2m-1\right) \left( \mathcal{G}-1\right) +\sum_{l=1}^{r}\lfloor
m\left( 1-\frac{1}{e_{l}}\right) \rfloor , & m>1 \\
&  \\
\mathcal{G}, & m=1 \\
\text{ }1, & m=0 \\
0, & m<0%
\end{array}%
\right.   \label{dim}
\end{equation}%
Here $\lfloor x\rfloor $ denotes the largest integer less or equal to $x$.

\subsection{Completeness theorem}

The next results (see the Appendix for proofs) provide necessary conditions
for the completeness of the discrete coherent states indexed by Fuchsian
groups.

\begin{theorem}
\textit{Let }$\{\left\vert z,B,n\right\rangle \}_{z\in \mathbb{C}^{+}}$%
\textit{\ be a system of coherent states attached to the }$nth$\textit{\
hyperbolic Landau level. If the subsystem }$\{\left\vert g\zeta
_{0},B,n\right\rangle \}_{g\in G}$\textit{\ indexed by the Fuchsian group }$%
G $\textit{\ associated with the }automorphic form $F_{0}$ of weight $m_{0}$%
, vanishing at one point $\zeta _{0}\in \mathbb{C}^{+}$ \textit{is complete,
then}
\begin{equation*}
m_{0}\geq \frac{1}{2}\frac{B-n}{1+n}\text{.}
\end{equation*}
\end{theorem}

If we can choose the automorphic form of weight $m_{0}=\frac{2\pi }{S_{G}}$,
where $S_{G}$ is the area of the fundamental domain the above theorem can be
rephrased as a necessary upper bound on $S_{G}$.

\begin{corollary}
\textit{Let }$\{\left\vert z,B,n\right\rangle \}_{z\in \mathbb{C}^{+}}$%
\textit{\ be a system of coherent states attached to the }$nth$\textit{\
hyperbolic Landau level. If the subsystem }$\{\left\vert g\zeta
_{0},B,n\right\rangle \}_{g\in G}$\textit{\ indexed by the Fuchsian group }$%
G $\textit{\ \ }vanishing at one point $\zeta _{0}\in \mathbb{C}^{+}$
\textit{is complete, then}
\begin{equation*}
S_{G}\leq 4\pi \frac{1+n}{B-n}\text{.}
\end{equation*}
\end{corollary}

Let's consider $\dim \mathfrak{H}ol_{G}^{m}\left( \mathbb{C}^{+}\right) \geq
2$. This guarantees the existence of an automorphic form of weight $m$
vanishing at a given $\zeta _{0}$, using appropriate linear combinations.
When $\mathcal{G}=0$ and $m\geq 2$, $m_{0}$ can be evaluated explicitly in
terms of the signature $(0,r,e_{1},...e_{r})$ of the group Fuchsian group $G$%
.\ \

\begin{corollary}
\textit{Let }$G\ $be a group of signature $(0,r,e_{1},...e_{r})$, with $\dim
\mathfrak{H}ol_{G}^{m}\left( \mathbb{C}^{+}\right) \geq 2$ and $m\geq 2$.
\textit{If the subsystem }$\{\left\vert g\zeta _{0},B,n\right\rangle
\}_{g\in G}$\textit{\ indexed by the Fuchsian group }$G$\textit{\ is
complete, then}
\begin{equation*}
\sum_{l=1}^{r}\lfloor 1-\frac{1}{e_{l}}\rfloor -2\leq 2\frac{1+n}{B-n}
\end{equation*}%
In particular, if $G=$ $PSL\left( 2,\mathbb{Z}\right) $, then
\begin{equation*}
\frac{1}{6}\geq 2\frac{1+n}{B-n}\text{.}
\end{equation*}
\end{corollary}

\textbf{Remark}. If we impose the frame condition on the coherent states, the
inequality
\begin{equation*}
m_{0}\geq \frac{1}{2}\frac{B-n}{1+n}\text{.}
\end{equation*}
 is an obvious consequence of Theorem
1 because the frame property is stronger than the completeness property. In the case of the Fuchsian group of dilations, it is possible to  use a standard perturbation argument
from wavelet theory \cite{AscensiBruna}, which assures that small
pseudohyperbolic perturbations of the index set of a wavelet frame keep the
wavelet frame property and obtain a strict inequality (this has been done in \cite{Seip} and \cite{WavFram}). However, it is not clear if  such a perturbation argument can be adapted to the case of a general Fuchsian group. We leave the problem as a question for the interested reader.

\section{Conclusion}

We have constructed discrete coherent states associated with the evolution of a particle under the action of a constant magnetic field when higher Landau levels are formed, first in the Euclidean and in the hyperbolic model.
Both in the higher Euclidean and the hyperbolic Landau levels, one can construct discrete coherent states by indexing the continuous ones by the discrete subgroups that reflect the symmetries of the underlying geometry.
The main conclusion is that, in both cases, the completeness of the coherent states depend explicitly on the size of the fundamental domain, on the order of the Landau Level
and on the intensity of the magnetic field. The analysis of the hyperbolic case is based on the properties of the automorphic form of weight $m$ associated with the Fuchsian group $G$  of the hyperbolic plane. If $G$ admits an automorphic form of weight  with a single zero inside $D$, then
$S_{G}=\frac{2\pi }{m}$ and we can choose the automorphic form of weight $m_{0}=\frac{\pi }{2S_{G}}$, where $S_{G}$ is the area of the fundamental domain. Then, the following restriction must be imposed for the completeness of the coherent states:
 \begin{equation*}
m_{0}\geq \frac{1}{2}\frac{B-n}{1+n}\text{.}
\end{equation*}

In terms of the area $S_{G}$ of the fundamental domain
\begin{equation*}
S_{G}\leq 4\pi \frac{1+n}{B-n}\text{.}
\end{equation*}

The methods used in this paper have their origins in several areas of mathematics, physics and signal analysis. It is not surprising that signal analysis and physics are strongly interrelated, since time-frequency (Gabor) analysis is the counterpart of the standard coherent states and time-scale (wavelet) analysis is the counterpart of affine coherent states and affine integral quantization \cite{BG}. But the arithmetic aspects connected to the hyperbolic geometry seem to have been somehow overlooked. Among the possible subgroups, only the Fuchsian group of dilations has been used in signal analysis \cite{Seip}, leading to the standard discretization of the half plane used in wavelet theory. We speculate that the discrete coherent states introduced in this paper may be useful in the analysis of signals, due to the variety of the discrete groups of the upper half-plane.  Finally, we would subscribe to the last sentence of the conclusion of \cite{BG}, since we believe it also applies to the current research:
\
\emph{(...) mutual irrigations between quantum physics and signal analysis deserve a lot more attention in future investigations.}

\begin{center}
\textbf{Appendix}
\end{center}

\textbf{Proof of Theorem 2. }Let $f\in L^{2}(\mathbb{R}^{+}, t^{-1}dt)$. Then
we can use Proposition1
\begin{equation}
W_{B,n}\left[ f\right] \left( z\right) =\gamma _{B,n}^{-\frac{1}{2}%
}\sum\limits_{k=0}^{n}\frac{\left( 2i\right) ^{k}}{k!}\left(
\begin{array}{c}
2B-n-1 \\
n-k%
\end{array}%
\right) y^{B-n-\frac{1}{2}+k}F^{\left( k\right) }\left( z\right)
\label{wavelet}
\end{equation}%
where%
\begin{equation*}
F=Ber_{2(B-n)-1}\left[ u\right] \in A_{2(B-n)-1}\left( \mathbb{C}^{+}\right)
\text{.}
\end{equation*}%
The idea of the proof is the following. Using the theory of automorphic
forms, we will construct a function $H\in A_{2(B-n)-1}\left( \mathbb{C}%
^{+}\right) $ vanishing at a point $\zeta _{0}\in \mathbb{C}^{+}$ and such
that, for $k=0,...n$, $H^{(k)}$ vanishes at $G\zeta _{0}$, the orbit of $%
\zeta _{0}$ under the action of $G$. Then set $F=H$ in (\ref{wavelet}); the
surjectivity of $Ber_{2(B-n)-1}$ assures the existence of $f\in L^{2}(%
\mathbb{R}^{+},t^{-1}dt)$ such that%
\begin{equation}
\left\{
\begin{array}{c}
H=Ber_{2(B-n)-1}\left[ f\right] \\
W_{B,n}\left[ f\right] \left( z\right) =0\text{, if }z\in G\zeta _{0}%
\end{array}%
\right. \text{;}  \label{H}
\end{equation}%
The function $H$ is constructed as follows. Let $F_{m_{0}}$ be a modular
form of weight $m_{0}$, that is, a function analytic on the upper-half plane
such that
\begin{equation}
F_{m_{0}}(z)=(cz+d)^{-2m_{0}}F_{m_{0}}\left( \frac{az+b}{cz+d}\right) \text{.%
}  \label{func}
\end{equation}%
If $G$ admits an automorphic form $F_{m_{0}}(z)$ vanishing at possible cusps
and vanishing at a point $\zeta _{0}\in \mathbb{C}^{+}$, the functional
equation (\ref{func}) implies that $F_{m_{0}}(z)$ vanishes at $G\zeta _{0}$.
Since
\begin{equation*}
\left( \Im z\right) ^{-1}\left\vert \Im\left( \frac{az+b}{cz+d}%
\right) \right\vert =\left\vert cz+d\right\vert ^{-2}\text{,}
\end{equation*}%
we have
\begin{equation*}
\left\vert F_{m_{0}}(z)\right\vert =\left( \Im z\right)
^{-m_{0}}\left\vert \Im \left( \frac{az+b}{cz+d}\right) \right\vert
^{m_{0}}\left\vert F_{m_{0}}\left( \frac{az+b}{cz+d}\right) \right\vert
\text{.}
\end{equation*}%
Thus, the function%
\begin{equation*}
\left( \Im z\right) ^{m_{0}}\left\vert F_{m_{0}}(z)\right\vert
=\left\vert \Im \left( \frac{az+b}{cz+d}\right) \right\vert
^{m_{0}}\left\vert F_{m_{0}}\left( \frac{az+b}{cz+d}\right) \right\vert
\end{equation*}%
is non-negative and continuous in the fundamental region $D$. Moreover, it
tends to $0$ as $\Im z\rightarrow \infty $ (this follows from an
argument using $q$-expansions \cite[pg. 94, formula (40)]{Serre}). Hence,
due to its $G$-invariance, it is bounded in the whole upper half-plane $%
\mathbb{C}^{+}$. As a result, the\ automorphic form $F_{m_{0}}(z)$ satisfies
\begin{equation}
\left\vert F_{m_{0}}(z)\right\vert \lesssim \left\vert \Im z\right\vert
^{-m_{0}}\text{, for every }z\in \mathbb{C}^{+}\text{.}  \label{3.5.6}
\end{equation}%
The above argument is well known in number theory (for instance, it is an
important step in the proof of Hecke's bound on Fourier coefficients of cusp
forms \cite[pg. 94]{Serre}). Now we argue by contradiction, supposing that $%
2m_{0}<\frac{B-n}{1+n}$. This implies the existence of $\epsilon >0$ such
that $m_{0}(n+1)=\frac{\alpha +1-\epsilon }{2}$, $\alpha =2(B-n)-1$. Define
\begin{equation*}
H(z)=(z+i)^{-\epsilon }\left[ F_{m_{0}}(z)\right] ^{n+1}(z)
\end{equation*}%
and observe that $H\neq 0$ and that the derivatives $H^{(k)}(z)$ vanish at $%
G\zeta _{0}$. The estimate (\ref{3.5.6}) then yields
\begin{equation}
\left\vert H(z)\right\vert \lesssim \left\vert z+i\right\vert ^{-\epsilon }(%
\Im z)^{\text{ }-(n+1)m_{0}}=\left\vert z+i\right\vert ^{-\epsilon }(%
\Im z)^{-\frac{\alpha +1-\epsilon }{2}}\text{.}  \label{est1}
\end{equation}%
Now let $w\in \mathbb{D}$. With the change of variables $z=i\frac{w+1}{1-w}$
one can write the integral in the unit disk. The detailed calculation
follows
\begin{equation*}
z+i=\frac{2i}{1-w};\text{ \ \ }\Im z=\frac{(1-\left\vert w\right\vert
^{2})}{\left\vert 1-w\right\vert ^{2}};\text{ \ \ \ \ \ }\left( \Im %
z\right) ^{\alpha }d\mu ^{+}(z)=\frac{(1-\left\vert w\right\vert
^{2})^{\alpha }}{\left\vert 1-w\right\vert ^{2\alpha +2}}d\mu ^{D}(w)\text{,}
\end{equation*}%
where $d\mu ^{D}(w=x+iy\in \mathbb{D})=dxdy$ is the area measure in the unit
disc and $d\mu ^{+}(z)(z\in \mathbb{C}^{+})=d\left( \Re z\right)
d\left( \Im z\right) $ is the area measure in the upper-half plane.\
Thus (\ref{est1}) becomes
\begin{equation*}
\left\vert \frac{1}{(1-w)^{\alpha +1}}H\left( i\frac{w+1}{1-w}\right)
\right\vert \lesssim (1-\left\vert w\right\vert ^{2})^{-\frac{\alpha
+1-\epsilon }{2}}\text{.}
\end{equation*}%
Now, in order to show that $H\in A_{2(B-n)-1}\left( \mathbb{C}^{+}\right) $%
,\ \ the integral can be estimated as follows.
\begin{eqnarray*}
\int_{\mathbb{C}^{+}}\left\vert H(z)\right\vert ^{2}\left( \Im z\right)
^{\alpha }d\mu ^{+}(z) &=&\int_{\mathbb{D}}\left\vert \frac{1}{(1-w)^{\alpha
+1}}H(i\frac{w+1}{1-w})\right\vert ^{2}(1-\left\vert w\right\vert
^{2})^{\alpha }d\mu ^{D}(w) \\
&\lesssim &\int_{\mathbb{D}}(1-\left\vert w\right\vert ^{2})^{-\alpha
-1+\epsilon }(1-\left\vert w\right\vert ^{2})^{\alpha }d\mu ^{D}(w) \\
&=&\int_{\mathbb{D}}(1-\left\vert w\right\vert ^{2})^{-1+\epsilon }d\mu
^{D}(w)<\infty \text{.}
\end{eqnarray*}%
The last inequality can easily be verified directly by definition of area
measure or using the reproducing kernel equation for Bergman spaces in the
unit disc. Thus, $H(z)\in A_{\alpha =2(B-n)-1}(\mathbb{C}^{+})$ vanishes on $%
G\zeta _{0}$ together with its derivatives and $H(z)$ satisfies (\ref{H}).
This is enough to finish the proof, since the existence of a nonzero $f\in
L^{2}(\mathbb{R}^{+},t^{-1}dt)$ such that $W_{B,n}\left[ f\right] \left(
z\right) \ $vanishes on the whole orbit $G\zeta _{0}$ leads to%
\begin{equation*}
c_{B,n}^{-\frac{1}{2}}\int\limits_{\mathbb{R}^{+}}\overline{\left\langle
t\mid z,B,n\right\rangle }f\left( t\right) \frac{dt}{t}=W_{B,n}\left[ f%
\right] \left( z\right) =0\text{, }z\in G\zeta _{0}\text{,}
\end{equation*}%
and contradicts the hypothesis of $\left\{ \left\vert g\zeta
_{0},B,n\right\rangle \right\} _{g\in G}$ being complete. Thus, the
condition $2m_{0}<\frac{B-n}{1+n}$ does not hold. As a result one must have $%
m_{0}\geq \frac{1}{2}\frac{B-n}{1+n}$.

\textbf{Proof of Corollary 1 and 2}. If $\dim \mathfrak{H}ol_{G}^{m}\left(
\mathbb{C}^{+}\right) \geq 2$ one can find an automorphic form of weight $m$
vanishing at a given $\zeta _{0}$, using appropriate linear combinations.
Moreover, if $\mathcal{G}=0$ and $m\geq 2$,
\begin{equation}
\dim \mathfrak{H}ol_{G}^{m}\left( \mathbb{C}^{+}\right)
=1-2m+2\sum_{l=1}^{r}\lfloor m\left( 1-\frac{1}{e_{l}}\right) \rfloor \text{.%
}  \label{dim2}
\end{equation}%
Then, comparing (\ref{dim2}) with the formula (\ref{area}) for $S_{G}$\ and
using Poincar\'{e}%
%TCIMACRO{\U{b4}}%
%BeginExpansion
\'{}%
%EndExpansion
s formula (\ref{Poincare}), gives:
\begin{equation*}
N\geq \dim \mathfrak{H}ol_{G}^{m}\left( \mathbb{C}^{+}\right) -1\geq 1\text{,%
}
\end{equation*}%
since $\dim \mathfrak{H}ol_{G}^{m}\left( \mathbb{C}^{+}\right) \geq 2$.\
Thus, the quantity%
\begin{equation*}
N(m_{0})=\frac{m_{0}S_{G}}{2\pi }
\end{equation*}%
is minimized when $N(m_{0})=1$, leading to the explicit value of the least
weight $m_{0}$:%
\begin{equation*}
m_{0}=\frac{2\pi }{S_{G}}=\left[ \sum_{l=1}^{r}\lfloor 1-\frac{1}{e_{l}}%
\rfloor -2\right] ^{-1}\text{.}
\end{equation*}%
The statement for $G=$ $PSL\left( 2,\mathbb{Z}\right) $ can be obtained by
using its signature $\left( 0,3;2,3,\infty \right) $\ or by showing directly
that the area of the fundamental domain is $S_{G}=\frac{\pi }{3}$.

\textbf{Acknowledgement. }The authors thank the reviewers for the carefull reading that resulted in a significant improvement of the paper. We also thank Ana Margarida Melo and Filippo Viviani for
explaining the geometric aspects of the theory of lattices and for suggesting to us that Fuchsian groups should be the proper objects to discretize objects in the hyperbolic model and Yuri Neretin for pointing to the work of Perelomov. L.D.~Abreu and P. Balazs were supported by Austrian Science Foundation (FWF)
START-project FLAME ('Frames and Linear Operators for Acoustical Modeling
and Parameter Estimation'; Y 551-N13); M. de Gosson by FWF project number P 23902-N13; Z. Mouayn has been partially supported by FCT (Portugal), through
European program COMPETE/FEDER and by FCT project PTDC/MAT/114394/2009

\end{document}